# Radio lighting based on dynamic chaos generators

Alexander S. Dmitriev[1,2], Elena V. Efremova[1,2], Mark Yu. Gerasimov[1], Vadim V. Itskov[1,2]

[1]Kotelnikov Institute of Radio Engineering and Electronics of RAS, Moscow, Russia
[2]Moscow Institute of Physics and Technology (State University), Dolgoprudnyi, Moscow Reg., Russia
chaos@cplire.ru

A problem of lighting objects and surfaces with artificial sources of noncoherent microwave radiation with the aim to observe them using radiometric equipment is considered. Transmitters based on dynamic chaos generators are used as sources of noncoherent wideband microwave radiation. An experimental sample of such a device, i.e., a radio lighting lamp based on a chaos microgenerator and its performance are presented.

## Introduction

By radio lighting we will understand an artificially created local noise (noise-like) field of wideband (ultrawideband) radiation, noncoherent over space and time in radio or microwave range. Radio lighting is implemented using one or several devices of noncoherent emission. Hitting nearby surfaces and objects, microwave emission is partly absorbed, partly passes through, and is partly diffused. When propagating further, it carries information about the environment with which it interacts. In this sense, the situation is similar to a situation with conventional (visible) light; the only difference is a different frequency band and different laws of interaction with the environment. For conventional light, there is a wonderful observation instrument such as a human eye. To extract information about objects in the area of radio lighting, special sensors or systems of such sensors are necessary.

There is a similarity between radiolight and conventional visible light [1]. In both cases, radiation is noncoherent and wideband, which eliminates interference effects and reduces observation issues to estimation of power (and possibly, spectral, as in the case of color vision) parameters of the received signal. The main distinction between radiolight and conventional light is a huge difference of characteristic frequency values, about five orders of magnitude. This means essentially lower potential resolution of radiolight in comparison with conventional light. However, there is a number of situations in which such a resolution is either acceptable, or indifferent.

The present paper addresses creation of effective radio lighting sources. Reception of radiolight is discussed briefly, only to evaluate potential distance range of radio lighting device operation.

## 1. Sources of radio lighting

Observing objects using noncoherent microwaves and other noncoherent signals in frequency ranges different from visible light is fruitfully used for many years in radio astronomy, in Earth observation from space [2-4] and in medical diagnostics [5, 6]. In these cases, noncoherent microwave emission is used, generated by natural processes such as intrinsic thermal emission of physical bodies in microwave range or diffraction of microwaves generated by powerful natural sources, e.g., the Sun. It gives a lot of information that can help us answer a number of basic issues related to radio lighting and its properties. However, the idea of radio lighting using local artificial sources similar to lighting devices of the visible range of the electromagnetic spectrum is mentioned in the literature as somewhat exotic (e.g., [1]).

One of the reasons for this is the lack of effective artificial sources of noncoherent microwave emission that might be used for radio lighting. Indeed, such sources must emit noise or noise-like wideband noncoherent signals, quite powerful in comparison with thermal emission. The devices must be as simple in operation as conventional lighting sources, such as incandescent, fluorescent, or LED lamps. Otherwise, radio lighting devices will be considered only in relation to special research equipment.

In microwave technology, two types of noise sources are used: gas-discharge tubes and semiconductor *pn*-diodes in the avalanche breakdown mode. Their main parameter is excess noise ratio (ENR), which is defined as the ratio of generated noise power to noise power of the resistor matched with a specified transmission line at the environment temperature. ENR is measured in dB. Gas-discharge tubes have typical values of ENR ~ 15 dB, that is approximately 30 times higher than the thermal noise power created by a matched resistor at the environment temperature 290 K. Thus, the tube generates noise that corresponds to approximately $9 \cdot 10^3$ K. ENR value of diode noise sources is ~ 30 dB (e.g., [7]), and their noise temperature is about $3 \cdot 10^5$ K, which corresponds to power spectral density $p \approx 4 \cdot 10^{-9}$ mW/MHz (–84 dBm/MHz). Integral power of the device in 1 GHz bandwidth is $4 \cdot 10^{-6}$ mW. Higher ENR can be obtained using amplifiers. However, essential amplification of the output power requires rather complex and expensive circuits [8].

In the following Sections, we will show that the reasonable value of the power of a standalone artificial radio lighting source is 0.1–1.0 mW. Thus, neither of abovementioned types of noise sources could be considered a suitable solution for radio lighting.

## 2. Dynamic chaos microwave transmitter as an artificial source of radio lighting

For radio lighting, we propose to use dynamic chaos generators [9] that might be considered sources of noise-like analog signals of the required frequency range.

These devices were developed over a number of years: from vacuum tubes in which natural delay of the distributed systems is used [10-12], to semiconductor devices with distributed oscillation systems on microstrips, to semiconductor devices with lumped-element oscillation systems [13]. These devices generate chaotic signals in the prescribed RF or microwave band.

Main stages of research and development of miniature microwave chaos generators since 2000 are shown in Fig. 1. Earlier, microwave chaos generators on active semiconductor elements existed only in the form of microstrip devices. By 2008, sufficient progress was achieved in designing lumped-element chaos generators, that gave more simple design, smaller size, better produceability, and lower cost. Additionally, such devices could be produced then in batches of hundred and thousand pieces.

A typical microwave lumped-circuit chaos generator is implemented on the basis of an oscillator with 2.5 freedom degrees, with a bipolar transistor as active element (Fig. 2). Mathematical model of the generator is a system of five first-order nonlinear ordinary differential equations:

$$\begin{aligned}
C_0 \dot{V}_{CE} &= I_1 - I_C, \\
L_1 \dot{I}_1 &= V_1 - V_{CE} - R_1 I_1, \\
C_1 \dot{V}_1 &= I_2 - I_1, \qquad (1) \\
L_2 \dot{I}_2 &= V_{BE} - V_1 - R_2 I_2, \\
C_2 \dot{V}_{BE} &= (V_E - V_{BE})/R_E - I_2 - I_B,
\end{aligned}$$

where $V_{CE}$, $V_{BE}$ are collector-emitter and base-emitter voltages, $V_1$ is voltage over capacitance $C_1$; $V_E$ is the source voltage; $I_1$, $I_2$, $I_C$, $I_B$ are currents through inductances $L_1$ and $L_2$, collector C and base B of the transistor, respectively. The structure of this oscillator is such that oscillations are excited in a certain frequency range. Generation in a prescribed frequency band and chaotic nature of oscillations are achieved by means of a proper setting of the oscillator parameters [13]. On the basis of the developed microwave chaos generators, ultrawideband (UWB) transceivers for wireless communications and wireless sensor systems [14] were designed.

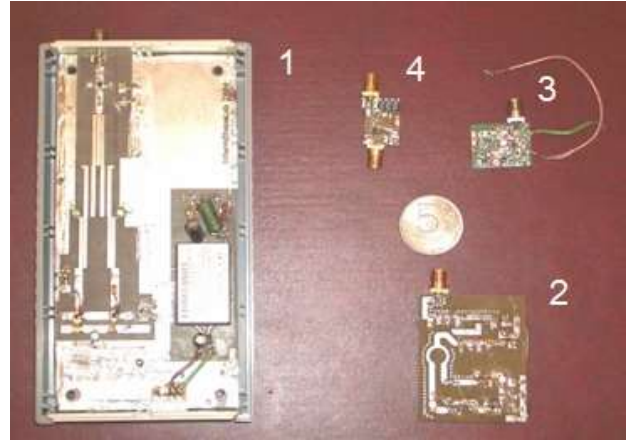

Fig. 1. Evolution of microwave chaos generators: (1) microstrip technology generator, 2003; (2) generator on three amplifier microchips with elements of microstrips, 2005; (3) lumped-element generator on three amplifier microchips; (4) lumped-element generator on a single bipolar transistor.

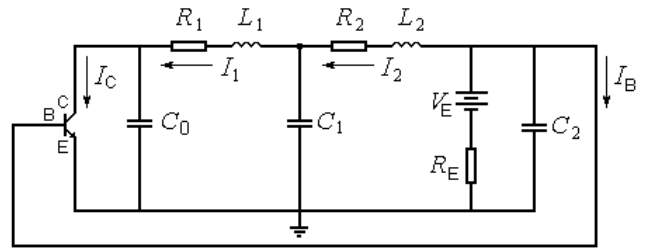

Fig. 2. Oscillator with 2.5 degrees of freedom.

The next stage of miniaturizing chaos generators and increasing their produceability is the design of integrated circuits (IC). The work in this direction was initiated several years ago [15, 16]. The IC design is based on oscillator (1) with several additional elements (Fig. 3). Typical topology of an IC chaos source with a single active element is shown in Fig. 4. So far, experimental samples of chaos generators are made with integral emission power of about 300μW in 3…7GHz frequency range (Fig. 5a). The devices are manufactured on silicon-germanium (SiGe) 0.25μm technology, the chip area is 1.6 sq.mm.

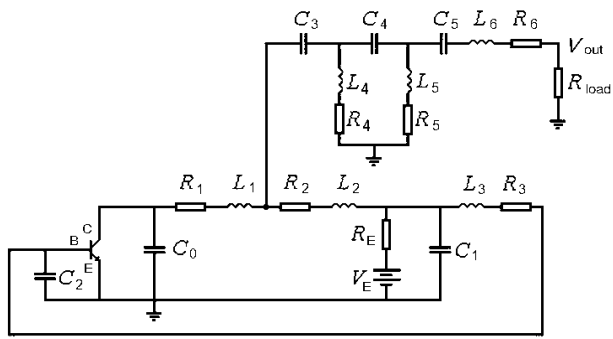

Fig. 3. Schematics of IC generator.

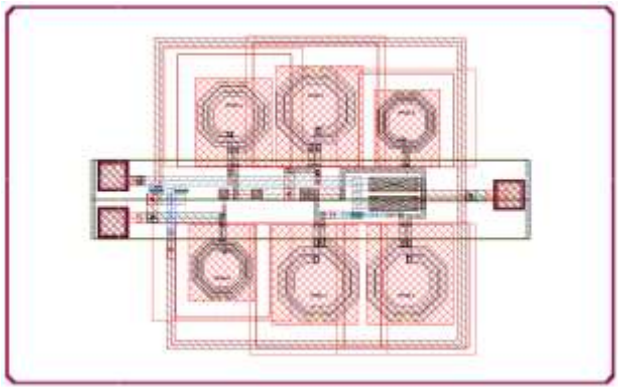

Fig. 4. IC topology.

Sources with such output power might be quite effectively used in the radio lighting devices.

Creation of dynamic chaos generator ICs drastically changes situation in the production area: now rather powerful sources of noncoherent UWB noise-like signals of microwave range can be produced at less cost and in large batches. As a result, there appears a new application area, namely, radio lighting.

The nearest counterparts of microwave UWB noncoherent sources in the frequency range visible to human eye are white LEDs. To obtain white light, a blue LED is used, emission of which is directed to yellow-emitting phosphor (Fig. 6a). White spectrum light is obtained by means of mixing the blue light from the blue LED with a longer-wavelength light from the phosphor (Fig. 6b). As a result, white LED emits noncoherent noise signal of $\Delta\lambda = 450…650$nm wavelength range. From the viewpoint of radiophysics, the white light bandwidth is ultra-wide, since $\Delta f/f = \Delta\lambda/\lambda > 0.25$, where $\Delta f$ is the difference of the upper and lower frequencies of the emission spectrum, and $f$ and $\lambda$ are average frequency and average wavelength, respectively.

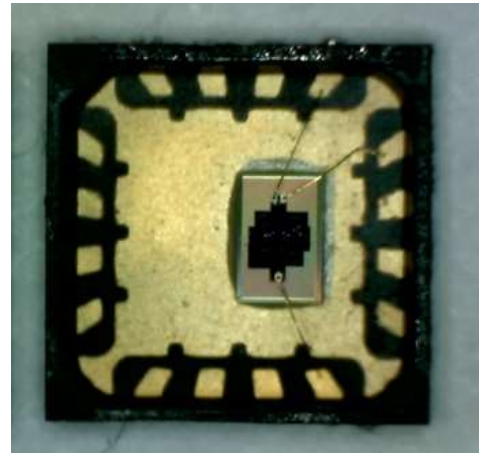
(a)

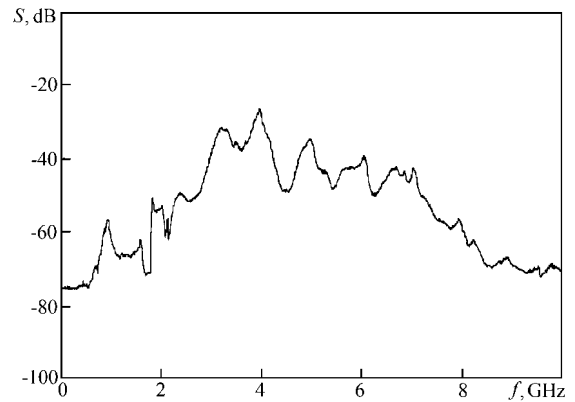
(b)

Fig. 5. Microgenerator of UWB chaotic oscillations: (a) packaged SiGe chip; (b) power spectrum.

Similarity of chaos microgenerators as sources of radio lighting and white-emitting LEDs becomes apparent, if we compare the spectrum of the LED (Fig. 6b) with that of the microgenerator (Fig. 5b – experiment, Fig. 7a – modeling). Such characteristics as waveform (Fig. 7b), rapidly decreasing autocorrelation function (Fig. 7c), and statistic distribution of instantaneous signal values which is close to Gaussian (Fig. 7d), confirm that the chaotic signal possess a set of features that provide noncoherent lighting of microwave frequency range.

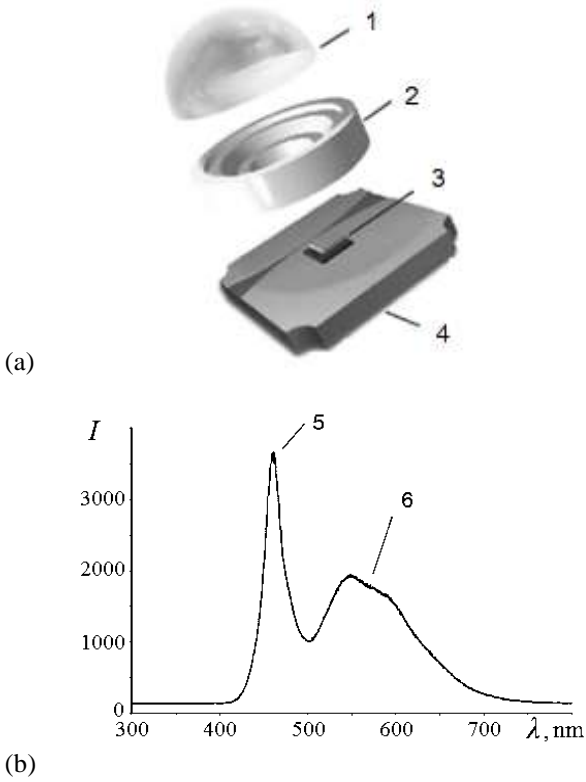

Fig. 6. White LED, (a) layout: (1) lens, (2) reflector, (3) LED matrix and phosphor, (4) substrate; (b) emission spectrum: (5) blue LED spectral line, (6) yellow phosphor fluorescence region.

An important common feature of microwave chaos generators and LEDs is noncoherence of the generated signal. This feature is necessary in order to use chaos microgenerators as lighting sources, since it provides uniform (without interference) coverage of the environment, as for a solitary or for multiple sources. The other important common property of LEDs and chaos emitting chips (CEC) is similarity of electric characteristics: both devices are low-voltage and can be used one at a time or in serial or parallel assemblies, e.g., in order to increase the power or to distribute emission over the space.

*Thus, dynamic chaos generators are effective sources of wideband noncoherent emission of microwave frequency range, likewise white LEDs are effective sources of wideband noncoherent emission of visible light spectrum.*

### 3. Models of microwave band lighting

Let us consider application principles of local compact sources of radio lighting based on UWB dynamic chaos generators. The situation here is essentially different from the situation with lighting in visible light range, which is associated with both physical characteristics and issues of scaling.

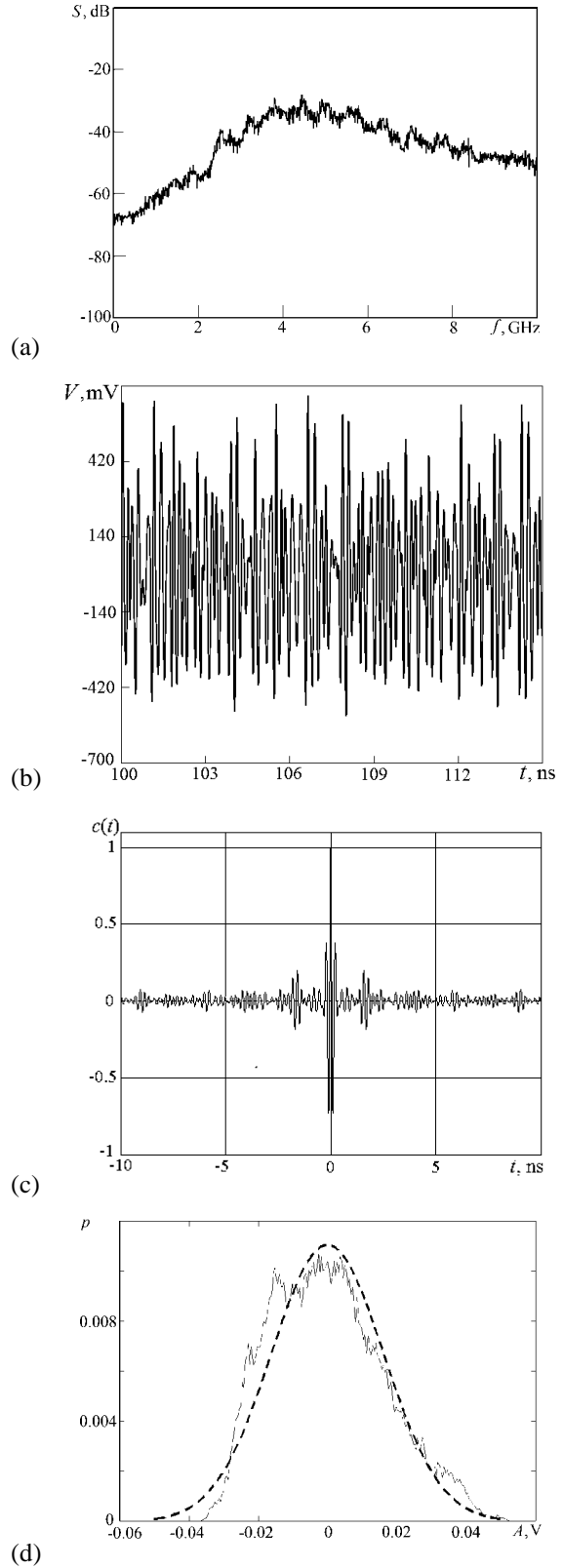

Fig. 7. Simulation of the signal at the output of chaos microgenerator, (a) waveform, (b) power spectrum; (c) autocorrelation function; (d) distribution $p$ of instantaneous values of signal amplitude $A$ at the antenna output (solid line – modeling, dashed line – Gaussian distribution.)

Consider typical geometrical configurations of sources, objects and the observation device. In Fig. 8a, an example of a "scene" with radio lighting sources *1*, *2*, *3*, an object that diffuses radio emission and allows it through *4*, and an observation device *5* are shown. In this example, the observation device receives emission directly from sources *1* and *2*; emission of sources *1* and *2* diffused by object *4*, and emission from object *3*, that passed through the object.

Judging from the relative position of radio lighting sources, objects, the observation device, average emission wavelength $\lambda$, and electric sizes of antennas of radio lighting sources and of the observation device, we can distinguish several typical "scenes."

Let us make the following assumptions that do not affect generality of the problem but essentially narrow the scope of considered models. Namely, we will consider only isotropic radio lighting sources (equally emitting in all directions) and consider the problem to be scalar (polarization issues will be ignored).

Let us divide the entire interval of distances $R$ from the observation device antenna to the "scene" (fragment of the environment) into two zones, judging from the average emission wavelength $\lambda$ and characteristic linear antenna dimension $L$ [17]:

Near field region and intermediate region (Fresnel region):

$$R \leq \frac{2L^2}{\lambda}; \qquad (2)$$

Far field region or radiation region (Fraunhofer region):

$$R > \frac{2L^2}{\lambda}. \qquad (3)$$

Note that for an isotropic radiator, the relation between directive gain $D = 1$ and the effective radiation area $S$ is defined with equation:

$$D = \frac{4\pi S}{\lambda^2} = 1 \Rightarrow S = \frac{\lambda^2}{4\pi}. \qquad (4)$$

In the further estimates, observation antenna aperture is assumed to have round shape. In this case, $S = \frac{\pi L^2}{4}$ and Eq. (4) gives $L = \frac{\lambda}{\pi}$.

For an observation device with isotropic antenna, only the far field (Fraunhofer) region is of practical interest:

$$R > \frac{2\lambda}{\pi^2} \approx \frac{\lambda}{5}. \qquad (5)$$

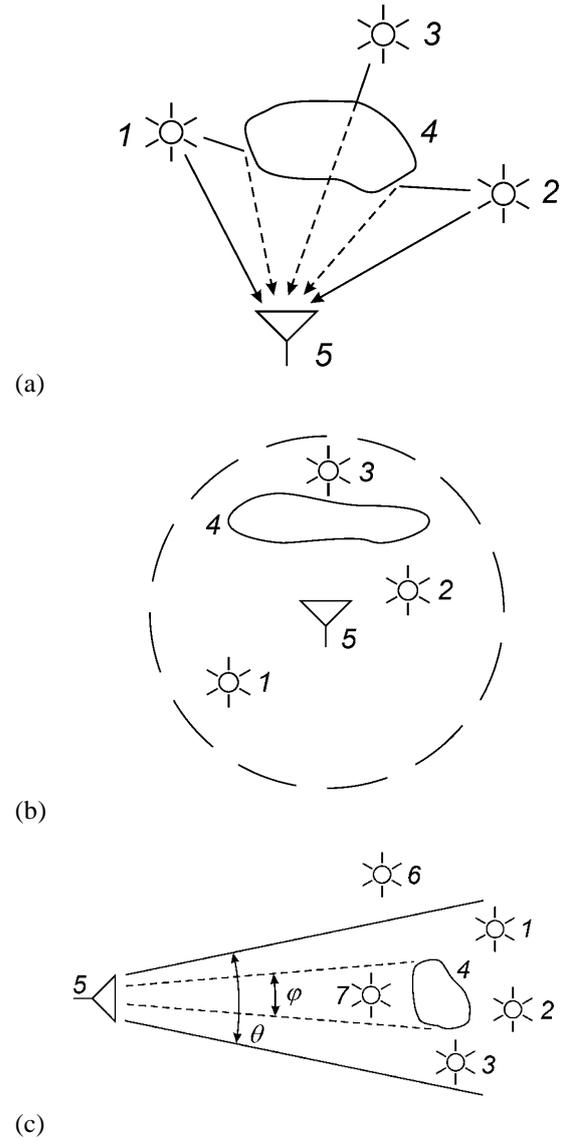

(a)

(b)

(c)

Fig. 8. Typical arrangements of sources, irradiated objects and observation device. *1, 2, 3, 6,* and *7* – radio lighting sources, *4* – object that diffuses and passes emission through, *5* – observation device. (a) irradiated "scene" with radio lighting sources; (b) second basic lighting model; (c) third basic lighting model; $\theta$ – width of observation device antenna pattern; $\varphi$ – angle by which an irradiated "scene" is seen from the measurement point.

In typical cases, all the sources are in the far field region (5) of the observation device antenna, and the waves that arrived at the observation device antenna are summed by power. In fact, the situation might be considered as point-to-point transmission, in which the observation device receives emission from some average source. The power of this average source can be estimated as the sum power of all the sources, considering them point-like and located approximately at equal distance from the measurement device. If the distances to the emission sources are essentially different, the powers of different sources must be taken into account regarding the corresponding distances. What is important in all of these cases, the waves from all the sources are noncoherent and are summed by power.

For a basic model of lighting in the considered case (the first basic model) we can take a model with a single radio lighting source (without diffusing or refracting objects), the emission of which comes directly to the observation device antenna. If the source power is known, then from its share that arrived to the observation device, one can estimate the distance to the source.

Let the observation device antenna have characteristic dimension $L$. Then the width of its directivity pattern in far field region is equal to $\theta \sim \lambda/L$, and the size of the resolution element is $\Delta \sim \dfrac{\lambda}{L} R$. As follows from condition (3), the distance to far zone border rapidly increases with an increase of the antenna size. For example, with $L = 5\lambda$ we have $R > 50\lambda$. If $\lambda = 7.5$ cm, then $R > 3.75$ m. If $L = 10\lambda$, then the distance to the far field region is increased up to $200\lambda$, or 15 m. The linear size of the antenna is 37.5 cm in the first case and 75 cm in the second case. With such reasonable antenna sizes, observations in the Fresnel zone can be provided for small and average size premises: up to 40 m$^2$ in the first case and 160 m$^2$ in the second case. Angular resolutions of the antennas are equal to 12° and 6° in the first and second cases, which gives space resolution $\delta_1 = \theta_1 R = 75$ cm and $\delta_2 = \theta_2 R = 37.5$ cm at the boundary of the Fraunhofer region for the antenna size $5\lambda$ and $10\lambda$, respectively. Within the Fresnel region, resolution is improved with decreasing distance down to the near field region, reaching the limit value $\sim \lambda$ at the distance of the order of the antenna size. Thus, in the case of lighting the spaces from tens to thousands of square meters using the discussed system characteristics, the area of the observed "scene" essentially exceeds the size of the resolution element, which means that power hitting the resolution element does not depend on the distance (for equal lighting.)

The second basic model of radio lighting describes the case in which the "scene" consists of surfaces (reflecting, refracting), which sizes essentially exceed the observation device antenna resolution element. Moreover, there may be radio lighting sources on the "scene," emission of which gets to the observation device antenna (Fig. 8b.)

The third lighting model is an intermediate case between the first and the second models. In this case, angular dimensions, relative to the measurement point, of the whole irradiated "scene" together with radio lighting sources are less than the width of the directivity pattern. This situation is shown in Fig. 8b, where $\theta$ is the width of the observation device antenna directivity pattern; $\varphi$ is the angle by which the irradiated "scene" is seen from the measurement point. Note that in this configuration, radio lighting sources may be placed within the sector corresponding to the directivity pattern of the observation antenna, as well as outside it. This is important, because if emission of the radio lighting sources falls directly to the observation device antenna, then its power prevails.

## 4. Estimation of the detection range of radio lighting sources

A point-like radio lighting source is characterized with emitted power $P$ and power spectral density $S(f)$ within the emission bandwidth $\Delta f$.

The power diffused by irradiated objects is estimated with radiolocation equation for spaced transmitter and receiver. The transmitter is represented with a radio lighting source with isotropic directivity pattern, whereas the receiver with an observation device. For the absolute cross-section, we take the area of the object projection on the plane perpendicular to the observation device directivity pattern. In the case of several lighting sources, the power that arrives from them to the object is summed (emission is noncoherent.)

If the radio lighting source with power $P_{src}$ irradiates an object with cross-section $S_{obj}$ at a distance $R_1$, then (in case of no propagation losses) the radiowave power $P_{obj}$ reflected by the object is defined with equation:

$$P_{obj} = \dfrac{P_{src} S_{obj}}{4\pi R_1^2}. \qquad (6)$$

Since the area $S_{obj}$ characterizes an isotropically diffusing object, emission power arriving at the receiver input (observation device) $P_{odev}$ placed at distance $R_2$ from the diffusing object is equal to:

$$P_{odev} = \dfrac{P_{obj}}{4\pi R_2^2} S_{odev} = \dfrac{P_{src} G_{odev} S_{obj} \lambda^2}{(4\pi)^3 R_1^2 R_2^2}, \qquad (7)$$

where $G_{odev}$ is the observation device antenna gain.

The maximum distance at which an irradiated object still can be observed (at a fixed distance from the light source to the object) is defined by equation:

$$R_{\max} = \sqrt{\frac{P_{\text{src}} G_{\text{odev}} S_{\text{obj}} \lambda^2}{(4\pi)^3 (P_{\text{odev}})_{\min} R_1^2}}, \qquad (8)$$

where $(P_{\text{odev}})_{\min}$ is the power that characterizes the sensitivity threshold of the observation device.

It is interesting to compare the power received by the observation device directly from the radio lighting source and the power reflected by the object. The power diffused from the object is related with the source power with Eq. (6), i.e., these power values differ by a factor $\frac{S_{obj}}{4\pi R_1^2}$. For example, if the distance from the radio lighting source to the object is 5 m and the absolute cross-section is equal to 1 m$^2$, then the reflected power is more than 300 times lower than the source power. At the same time, if the source evenly irradiates a 100-m$^2$ surface at the same distance, then the power of the twice-reflected signal is already ~ 0.3 of $P_{\text{src}}$. Thus, total reflected power can change in a wide range.

Hence, to estimate the maximum possible range of the "scene" observation, it is reasonable to use the maximum distance range of the radio lighting source, which can be considered as the upper limit of the possible observation range also for the other lighting models.

## 5. Observation device sensitivity

For the sensitivity threshold of devices, measuring intensity of fluctuating electromagnetic emission, the value of noise signal at the receiver input is taken, equal by the output amplitude to rms deviation of the fluctuating signal caused by inherent noises of the amplification path.

An optimum device to receive noise-like signals of various physical nature, including electromagnetic radiation, is a device (Fig. 9) composed of an ideal amplifier, a quadratic detector, and a low-pass filter that accumulates the signal [4].

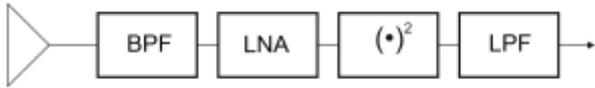

Fig. 9. Block diagram of the receiver: BPF – band-pass filter, LNA – low-noise amplifier, (.)$^2$ – quadratic detector, LPF – low-pass filter.

Sensitivity of such a radiometric receiver is equal to:

$$\Delta T = \sqrt{2} T_{\text{rec}} \sqrt{\frac{\Delta F}{\Delta f}}, \qquad (9)$$

where $T_{\text{rec}}$ is receiver temperature, $\Delta f$ is receiver input bandwidth, $\Delta F$ is receiver output bandwidth.

Let us estimate the distance at which radio lighting source can be detected using such a receiver. We use a standard procedure of calculating radio link budget $M$ in dB [17]:

$$M = P_{\text{src}} + G_{\text{src}} + G_{\text{odev}} - \left(\frac{E_b}{N_0}\right)_{\text{req}} - S - (kT) - L_R - L_0, \qquad (10)$$

where $G_{\text{src}}$ [dB] is the source antenna gain; $(E_b/N_0)_{\text{req}}$ [dB] is the required energy-per-bit to noise-spectral-density ratio in dB; $S$ [dB/s] is data transmission rate in dB per second; $kT$ [dBm] is spectral power of thermal noise at temperature $T$, $k$ is Boltzmann constant; $L_0$ [dB] is the loss due to propagation from the radio lighting source to the observation device at 1m distance; $L_R$ [dB] is additional attenuation of the received signal when the distance is increased from 1 m to $R$ m.

In our case, $(E_b/N_0)_{\text{req}}$ can be obtained from signal-to-noise ratio:

$$\frac{P_{\text{rec}}}{N} = \frac{E_b R}{N_0 W}. \qquad (11)$$

The threshold receiver power is equal to:

$$P_{\text{rec}} = k \Delta T \Delta f = k\sqrt{2} T_{\text{rec}} \sqrt{\frac{\Delta F}{\Delta f}} \Delta f. \qquad (12)$$

$E_b/N_0$ ratio can be obtained from Eq. (11):

$$\frac{E_b}{N_0} = \frac{P_{\text{rec}} W}{NR} = \frac{P_{\text{rec}}}{N_0 R}, \qquad (13)$$

hence:

$$\frac{E_b}{N_0} = \frac{k\sqrt{2} T_{\text{rec}} \sqrt{\frac{\Delta F}{\Delta f}} \Delta f}{N_0 R} = \frac{\sqrt{2}\sqrt{\frac{\Delta F}{\Delta f}} \Delta f}{R} = \sqrt{2}\sqrt{\frac{\Delta f}{\Delta F}}. \qquad (14)$$

With the receiver input amplifier bandwidth equal to $\Delta f$ =1 GHz and low-pass filter bandwidth $\Delta F$ =10 Hz, $(E_b/N_0)_{\text{req}}$ is equal to 41.5 dB.

Assuming $G_{\text{src}} = G_{\text{odev}} = 0$ dB, $(E_b/N_0)_{\text{req}}$ = 41.5 dB, transmission rate inversely proportional to accumulation time, 10 bps, $kT = -174$ dBm, link margin M = 10 dB and $L_0 = 44$ dB (for the average wavelength 7.5 cm), from equation (10) we obtain $L_R = 68.5$ dB. This means, for example, that the radio lighting source with 1 mW emission power can be detected using an observation device with omnidirectional antenna from the distance above 2.5 km.

## 6. Radio lighting lamp

Similar to LED lamp, a radio lighting lamp (RLL) can be created with a chaos generator microchip as active element. The idea of such a lamp is as follows.

A dynamic chaos microgenerator together with an antenna make up a source of noise-like UWB microwave radiation (Fig. 10a). Such a source can be considered a prototype of the radio lighting lamp. This prototype is then structurally transformed both in shape and operation mode into a usual "bulb" with 220 VAC power supply.

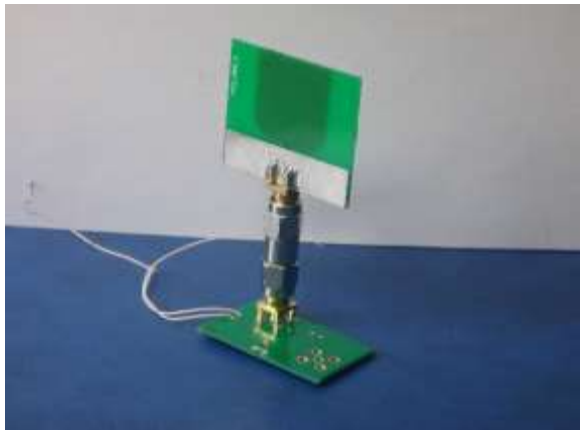

(a)

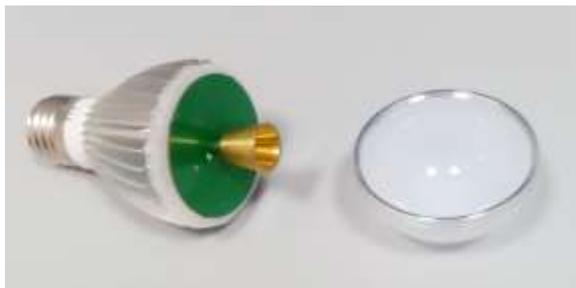

(b)

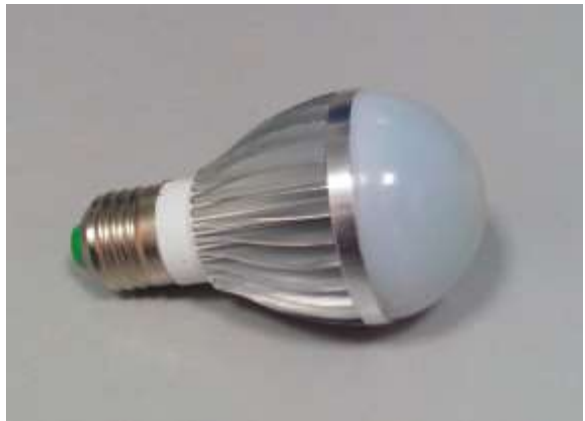

(c)

Fig. 10. Radio lighting bulb, (a) lamp prototype, a chaos generator with antenna attached; (b) emitter device with antenna (left) and case cover (right); (c) assembled device in the form of a standard lighting lamp.

An experimental sample of the radio lighting lamp RLL-2 was developed on the basis of dynamic chaos generator microchip that produces chaotic signal of 3…7 GHz frequency band (Section 3). The lamp emission power is equal to approx. 300 μW. The lamp contains a printed circuit board (PCB) with SMT electronic components and an antenna (Fig. 10b), and a secondary power supply that transforms 220 VAC to 5 VDC. Upper metal-coated PCB side together with a cone-like element form a discone emitting antenna [19, 20]. Additionally, a LED is placed on the upper PCB side to indicate the device mode (on/off). The lamp driver is packed in standard E27 socket case for LED lamps (Fig. 10c).

To deploy indoor or open-space radio lighting is as simple as to screw the bulb in a standard lantern receptacle connected to ordinary power line and to press the switch button.

Other designs of radio lighting sources can use various low-voltage DC sources, e.g., 12-V car lighter, 5V USB connectors of notebooks and tablets, etc.

## Conclusions

Development of efficient miniature sources of wideband noncoherent microwave emission on the basis of dynamic chaos gives us a way to create artificial radio lighting. With modern methods of receiving UWB signals we can develop a new class of equipment for wide-use info-communication infrastructure. Such an equipment can be used during the polar night in the Far North, to moor vessels in fog conditions, to navigate aircrafts by landing, etc.

Prototypes of devices for observing irradiated objects are traditional radiometer-type receivers [2-6] and log-detector sensors [14]. The latter have sufficiently high sensitivity and wide dynamic range even at room temperature. For instance, a receiver with a log detector and an omnidirectional antenna can sense the radio light from a 300 μW-emission power bulb at distances from tens of centimeters to several hundred meters.

**Acknowledgments** This study is supported by the Russian Science Foundation (project №16-19-00084).

## References

1. *Polivka J., Fiala P., Machac J.* Microwave noise field behaves like white light // Progress in Electromagnetics Research. 2011. V. 111. P. 311.

2. *Shutko A.M.* Microwave radiometry of water surfaces and soils. Moscow: Nauka, 1986 (in Russian).

3. *Armand N.A., Polyakov V.M.* Radio Propagation and Remote Sensing of the Environment. N.Y.: CRC Press, 2005.

4. *Sharkov E. A.* Radiothermal remote sensing of Earth: physical principles. Vol. 1. Moscow: SRI of RAS, 2014 (in Russian).

5. *Gulyaev Yu.V., Godik E.E.* Physical fields of biological objects // Herald of USSR Academy of Sciences. 1983. No. 8. p. 118 (in Russian).

6. *Gulyaev Yu.V.* Physical fields and human emission. New non-invasive methods of medical diagnostics. Moscow: S. Vavilov RBOF Znaniye, 2009 (in Russian).

7. NC100/200/300/400 Series Chips and Diodes // www.noisecom.com/products/components/nc100-200-300-400-series-chips-and-diodes.

8. *Bezrukov V.A.* Two-level noise generators TGN // Sovremennaya elektronika. 2011. No. 7. p. 28 (in Russian).

9. *Dmitriev A.S., Efremova E.V., Gerasimov M.Yu., Itskov V.V.* Radiolighting on the basis of ultrawideband generators of dynamic chaos // Radiotekhnika I elektronika. 2016. vol. 61. No. 11. P. 1073 (in Russian).

10. *Kislov V. Ya., Zalogin N. N., Myasin E. A.* Investigation of stochastic oscillation processes in time-delay generators // Radiotekhnika I elektronika. 1979. vol. 24. No. 6. P. 1118 (in Russian).

11. *Bezruchko B.P. , Kuznetsov S.P. , Trubetskov D.I.* Experimental observation of stochastic self-oscillations in the electron beam-backscattered electromagnetic wave dynamic system // JETF Letters. 1979. V. 29. N. 3. P. 162.

12. *Anisimova Yu. V., Dmitriev A. S., Zalogin N. N., Kalinin V. I., Kislov V. Ya., Panas A. I.* A mechanism for the transition to chaos in the system of an electron beam and an electromagnetic wave // JETF Letters. 1983. V. 37. N. 8. P. 458.

13. *Dmitriev A.S., Efremova E.V., Panas A.I., Maksimov N.A.* Generation of chaos. Moscow: Tekhnosfera, 2012, 424 p. (in Russian).

14. *Dmitriev A.S., Kletsov A.V., Laktyushkin A.M., Panas A.I., Starkov S.O.* Ultrawideband wireless communications based on dynamic chaos // Journal of communications technology and electronics. 2006. T. 51. № 10. C. 1126.

15. *Dmitriev A.S., Efremova E.V., Nikishov A.Yu.* Dynamic chaos in the microwave range generated in SiGe-based autooscillatory structure // Technical Physics Letters. 2009. T. 35. №. 12. C. 1090-1092.

16. *Dmitriev A.S., Efremova E.V., Nikishov A.Yu.* Generating dynamic microwave chaos in self-oscillating ring system based on complementary metal-oxide-semiconductor structure // Technical Physics Letters. 2010, vol. 36, №5, pp. 430–432.

17. *Panchenko B.A.* Antennae. Moscow: Goryachaya liniya – Telekom, 2015 (in Russian).

18. *Sklar B.* Digital communications. Fundamentals and applications. 2nd edition. Prentice Hall, PTR. 2001.

19. *Kaloshin V.A., Martynov E.C., Skorodumova E.A.* Investigation of characteristics of polyconic antenna in wide frequency range // *Radiotekhnika I elektronika*. 2011. vol. 56. No. 9. P. 1094 (in Russian).

20. *Kaloshin V.A., Skorodumova E.A.* Disc-polyconic antenna // Antennas. 2011. no. 10. No. 173. P.79 (in Russian).